\documentclass[12pt,english]{article}
\pdfoutput=1
\usepackage[T1]{fontenc}
\usepackage{fullpage}
\usepackage{authblk}
\usepackage{graphicx,array}
\usepackage{cite}
\usepackage{color}
\usepackage{latexsym}
\usepackage{amsthm}
\usepackage{amsmath}
\usepackage[titletoc]{appendix}
\usepackage{enumitem}
\usepackage{amssymb}
\usepackage{color}
\usepackage[unicode=true,
 bookmarks=true,bookmarksnumbered=false,bookmarksopen=false,
 breaklinks=false,pdfborder={0 0 1},backref=false,colorlinks=true]
 {hyperref}
\hypersetup{pdftitle={Entanglement of Purification and Multiboundary Wormhole Geometries},
 pdfauthor={Ning Bao, Aidan Chatwin-Davies, and Grant N. Remmen},
 citecolor=black,linkcolor=black,urlcolor=black}
\usepackage{breakurl}
\usepackage[hang,flushmargin]{footmisc} 
\usepackage{listings}
\usepackage{mathrsfs}
\usepackage{amsthm}

\newcommand{\Eq}[1]{Eq.~(\ref{#1})}

\newcommand{\Sec}[1]{Sec.~\ref{#1}}

\newcommand{\Fig}[1]{Fig.~\ref{#1}}

\newcommand{\Ref}[1]{Ref.~\cite{#1}}
\newcommand{\Refs}[1]{Refs.~\cite{#1}}
\usepackage{accents}
\newcommand{\dbtilde}[1]{\accentset{\approx}{#1}}

\newcommand{\Hil}{\mathcal{H}}

\newcommand{\ket}[1]{| #1 \rangle}

\newcommand{\ketbra}[2]{|#1\rangle \langle #2 |}

\newcommand{\be}{\begin{equation}}
\newcommand{\ee}{\end{equation}}

\DeclareMathOperator{\Tr}{Tr}

\usepackage{xcolor}
\usepackage{color}

\begin{document}

\interfootnotelinepenalty=10000

\hfill

\vspace{2cm}
\thispagestyle{empty}
\begin{center}
{\LARGE \bf
Entanglement of Purification and Multiboundary Wormhole Geometries
}\\
\bigskip\vspace{1cm}{
{\large Ning Bao,${}^a$ Aidan Chatwin-Davies,${}^{b,c}$ and Grant N. Remmen${}^a$}
} \\[7mm]
 {\it ${}^a$Center for Theoretical Physics and Department of Physics \\
     University of California, Berkeley, CA 94720, USA and \\
     Lawrence Berkeley National Laboratory, Berkeley, CA 94720, USA \\[1.5mm]
${}^b$Walter Burke Institute for Theoretical Physics \\
    California Institute of Technology, Pasadena, CA 91125, USA\\[1.5mm]
 ${}^c$KU Leuven, Institute for Theoretical Physics\\Celestijnenlaan 200D B-3001 Leuven, Belgium} \let\thefootnote\relax\footnote{\noindent e-mail: \url{ningbao75@gmail.com}, \url{aechatwi@gmail.com}, \url{grant.remmen@berkeley.edu}} \\
 \end{center}
\bigskip
\centerline{\large\bf Abstract}
\begin{quote} \small
We posit a geometrical description of the entanglement of purification for subregions in a holographic CFT. The bulk description naturally generalizes the two-party case and leads to interesting inequalities among multi-party entanglements of purification that can be geometrically proven from the conjecture. Further, we study the relationship between holographic entanglements of purification in locally-AdS\textsubscript{3} spacetimes and entanglement entropies in multi-throated wormhole geometries constructed via quotienting by isometries. In particular, we derive new holographic inequalities for geometries that are locally AdS\textsubscript{3} relating entanglements of purification for subregions and entanglement entropies in the wormhole geometries.

\end{quote}
	
\setcounter{footnote}{0}

\newpage
\tableofcontents
\newpage

\section{Introduction}
One of the main areas of research in the modern AdS/CFT program is the relationship between holography and quantum information theory.
In particular, the most established of these ideas is the notion of holographic entanglement entropy, which associates the entanglement entropy of a boundary subregion of the conformal field theory (CFT) with the area of a minimal surface in the bulk asymptotically Anti-de Sitter (AdS) geometry \cite{Ryu:2006bv}.
Because the existence of the bulk geometry places constraints on the set of allowed entropies associated with the holographic quantum state (which are not satisfied by general quantum states) \cite{Hayden:2011ag, Bao:2015bfa}, these entanglement entropies can be used to characterize which states in which conformal field theories can be dual to classical bulk spacetimes in the large-$N$ limit.

A recent development is the conjecture that a related information-theoretic object, the entanglement of purification \cite{terhal2002entanglement}, is also dual to a bulk geometric object \cite{Takayanagi:2017knl, Nguyen:2017yqw}.
The conjecture was motivated by demonstrating that the bulk object obeys the same known set of inequalities obeyed by the entanglement of purification.
The entanglement of purification for a mixed state is interesting to study because it is an entanglement measure of the purification of a given mixed state that preserves the notion of the bipartition in the mixed state and extends this to the full purification.
This conjecture has also motivated the definitions of conditional \cite{Bao:2017nhh, Espindola:2018ozt} and multipartite \cite{Bao:2018gck, Umemoto:2018jpc} entanglements of purification and also their conjectured bulk duals, which are particular surfaces in the asymptotically-AdS geometry.

We will formulate a candidate geometric object dual to multipartite entanglement of purification for holographic states that are dual to three-dimensional asymptotically-AdS geometries. Using this conjecture, we will geometrically derive a class of new holographic inequalities relating different $n$-party entanglements of purification among each other. This is an additional constraint that holographic states must satisfy, akin to the holographic entropy cone \cite{Bao:2015bfa}, but instead relating entanglements of purification.

The bulk surfaces that are conjectured to be dual to entanglement of purification have been studied previously in a different context: the creation of holographic wormhole geometries via quotients of vacuum AdS\textsubscript{3} by isometries \cite{Brill:1995jv,Aminneborg:1997pz,Brill:1998pr,Balasubramanian:2014hda, Marolf:2015vma}.
We find that the entanglement of purification in vacuum AdS\textsubscript{3} appears to map into boundary-homologous minimal surfaces in the constructed wormhole geometries, which via the Ryu-Takayanagi formula compute entanglement entropies of reduced states on entire boundaries.
In this work, we will explore possible applications of this observation that advance the study of both AdS\textsubscript{3} wormhole geometries and holographic entanglement of purification more generally. Specifically, we will use the equivalence of these objects to compute new constraints relating entanglements of purification of the vacuum state and entanglement entropies of wormholes in AdS\textsubscript{3}. 
We will also use these tools to study potential consequences for Hilbert space factorization in the wormhole geometries resultant from the identification.

This work is organized as follows.
In \Sec{sec:EPandHolo}, we will review the definition and holographic interpretation of two-party entanglement of purification and state its conjectured multipartite generalization.
From this conjecture, we will then derive a set of inequalities that multipartite entanglements of purification must satisfy for holographic states.
In \Sec{sec:WHviaOrbifolds}, we will briefly review the method of constructing wormhole geometries in AdS\textsubscript{3} via identifications in the bulk.
We will synthesize the two narratives in \Sec{sec:interpretingEP} and derive new results relating the entanglement of purification and entanglement entropies in different holographic geometries. 
We conclude with a discussion of future directions in \Sec{sec:disc}.

\section{Entanglement of Purification and Holography}
\label{sec:EPandHolo}

Given a density matrix $\rho_{AB}$, the entanglement of purification is defined as 
\be 
E_P(A:B)=\inf_{A'B'} S(AA').
\ee
The infimum is taken over all choices of auxiliary Hilbert spaces $\Hil_{A^\prime}$ and $\Hil_{B^\prime}$ and over all purifications $\ket{\Psi}_{AA'BB'}$ on the total Hilbert space such that $\Tr_{A'B'} \ketbra{\Psi}{\Psi} = \rho_{AB}$.
The quantity $S(AA')$ denotes the Von Neumann entropy of the reduced state on $\Hil_{AA'}$, $\rho_{AA'} = \Tr_{BB'} \ketbra{\Psi}{\Psi}$.
The entanglement of purification is known to obey the following inequalities for all quantum systems \cite{terhal2002entanglement,Bagchi:2015}:
\be 
\begin{aligned}
\min (S(A),S(B))& \geq E_P(A:B)\geq \frac{1}{2}I(A:B) \\
E_P(A:BC)& \geq E_P(A:B) \\
E_P(A:BC)& \geq \frac{1}{2}I(A:B)+\frac{1}{2}I(A:C),
\end{aligned}
\ee
where $I(A:B)=S(A)+S(B)-S(AB)$ is the mutual information between the $A$ and $B$ subsystems.
Moreover, $E_P(A:B)+E_P(A:C)\geq E_P(A:BC)$ for pure $\rho_{ABC}$.

\begin{figure}[h]
\centering
\includegraphics[scale=0.45]{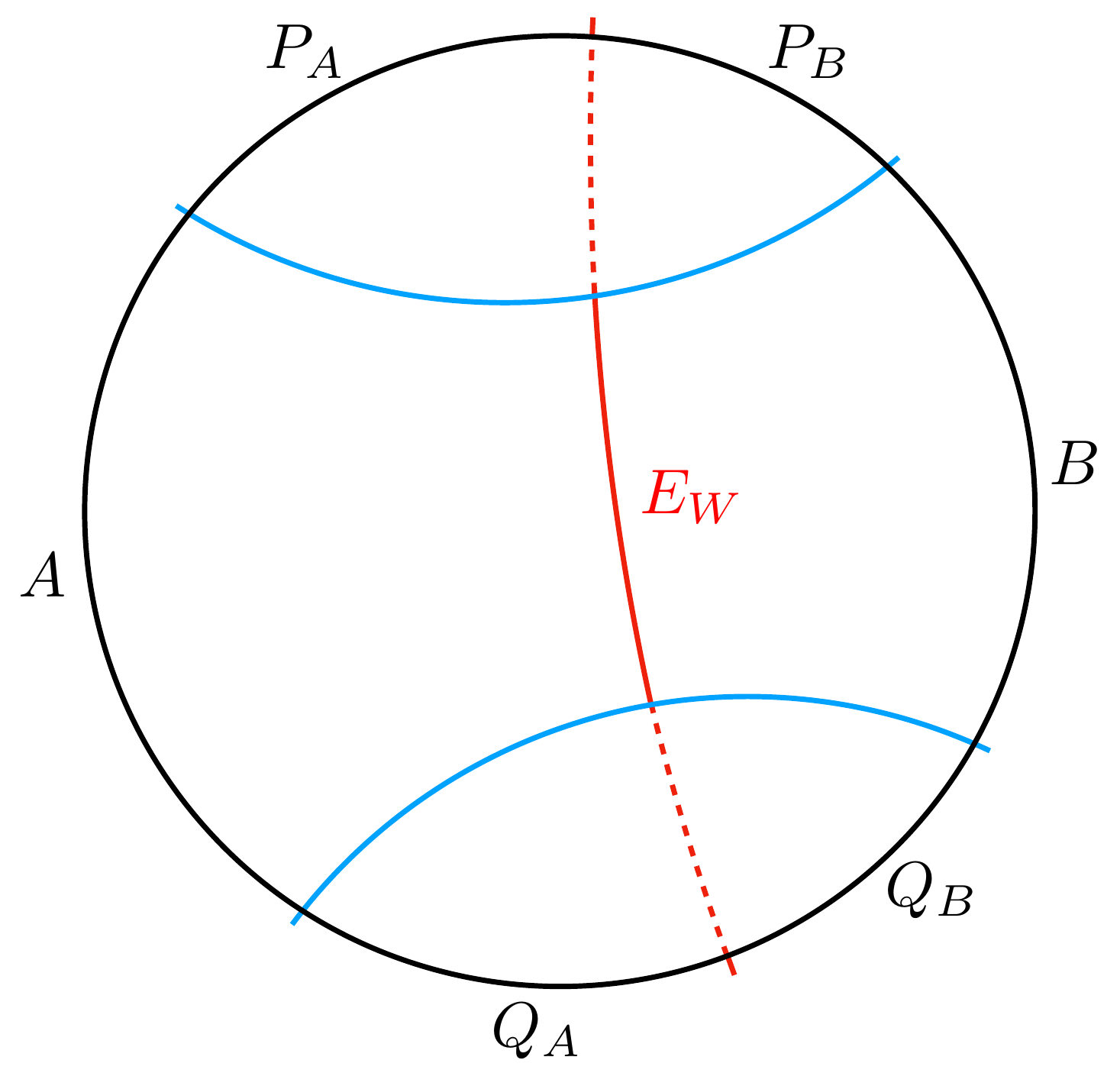}
\caption{The entanglement wedge cross section of the $AB$ boundary subsystem, with size $E_W$.
This object, depicted as a red line, is the minimal surface that totally partitions the entanglement wedge into a region adjacent to $A$ and one adjacent to $B$.
It can be extended to a boundary-anchored geodesic (red dashed line), which further partitions the complement of $A \cup B$ into $P_A$, $P_B$, $Q_A$, and $Q_B$.} 
\label{fig:2pep}
\end{figure}

The conjectured dual holographic object is the area, $E_W$, of the entanglement wedge cross section, which is defined as the minimal surface that partitions the $\rho_{AB}$ entanglement wedge into a region adjacent to only $A$ and one adjacent to only $B$; see \Fig{fig:2pep}. 
The entanglement wedge is defined on a spatial slice to be the bulk region enclosed by the union of the boundary subregions and the Ryu-Takayanagi surfaces homologous to them.
The conjecture is motivated by the fact that $E_W$ obeys the same inequalities as $E_P$, as shown in \Refs{Takayanagi:2017knl, Nguyen:2017yqw}.
For holographic CFTs, $E_W$ is conjectured to be calculable in terms of the entropies of $\rho_{AB}$ and its partial transpose \cite{Tamaoka:2018ned}.

Furthermore, there is an additional inequality that is true for $E_W$ that is not true for $E_P$ for generic quantum states:

\begin{equation}
E_W(AC:BD)\geq E_W(A:B)+E_W(C:D).
\end{equation}
This inequality could, as in the case of the holographic entanglement entropy inequalities, serve as another way to determine which quantum states have entropic properties consistent with being dual to classical bulk gravity solutions.

As a new result, we also note that any $E_W(A:B)$ for single intervals $A$ and $B$ can be extended to the boundary as a subset of a minimal surface computing a Ryu-Takayanagi entanglement entropy, generically of some region $AP_AQ_A$ with entanglement entropy $S(AP_AQ_A)$ and complement $BP_BQ_B$. Note that the $P$s and $Q$s are fixed by the choice of $A$ and $B$ and are thus not independent.
It is then also immediately true that 
\be 
S(AP_AQ_A)\geq E_W(A:B)+E_W(P_A:P_B)+E_W(Q_A:Q_B),
\ee 
as the $E_W$ are less constrained than the entanglement entropy.

\subsection{Multipartite entanglement of purification}

There is a multipartite generalization of the entanglement of purification. 
For clarity, we will first give the tripartite generalization:
\begin{equation}
E_P(A:B:C)=\inf_{A'B'C'} \frac{1}{3} \left[S(AA')+S(BB')+S(CC')\right] ,
\end{equation}
where the infimum is taken over all auxiliary Hilbert spaces $\Hil_{A'}$, $\Hil_{B'}$, $\Hil_{C'}$, and over all purifications $\ket{\Psi}_{AA'BB'CC'}$ of a given $\rho_{ABC}$. More generally, 
\be
\label{eq:muliepdef}
E_P (A_1:A_2:\ldots:A_n) = \inf_{A'_1 A'_2 \cdots A'_n} \frac{1}{n} \left[S(A_1 A'_1) + S(A_2 A'_2) + \cdots + S(A_n A'_n)\right], 
\ee
where $\ket{\Psi}_{A_1 A'_1 A_2 A'_2 \cdots A_n A'_n}$ is a purification of $\rho_{A_1 A_2 \cdots A_n}$.
It was shown in \Refs{Bao:2018gck, Umemoto:2018jpc} that this object obeys generalizations of the inequalities listed for bipartite $E_P$ for all quantum states, though those inequalities will not be used further in this work; we will rely only on the multipartite definition itself.

\begin{figure}
\centering
\includegraphics[scale=0.45]{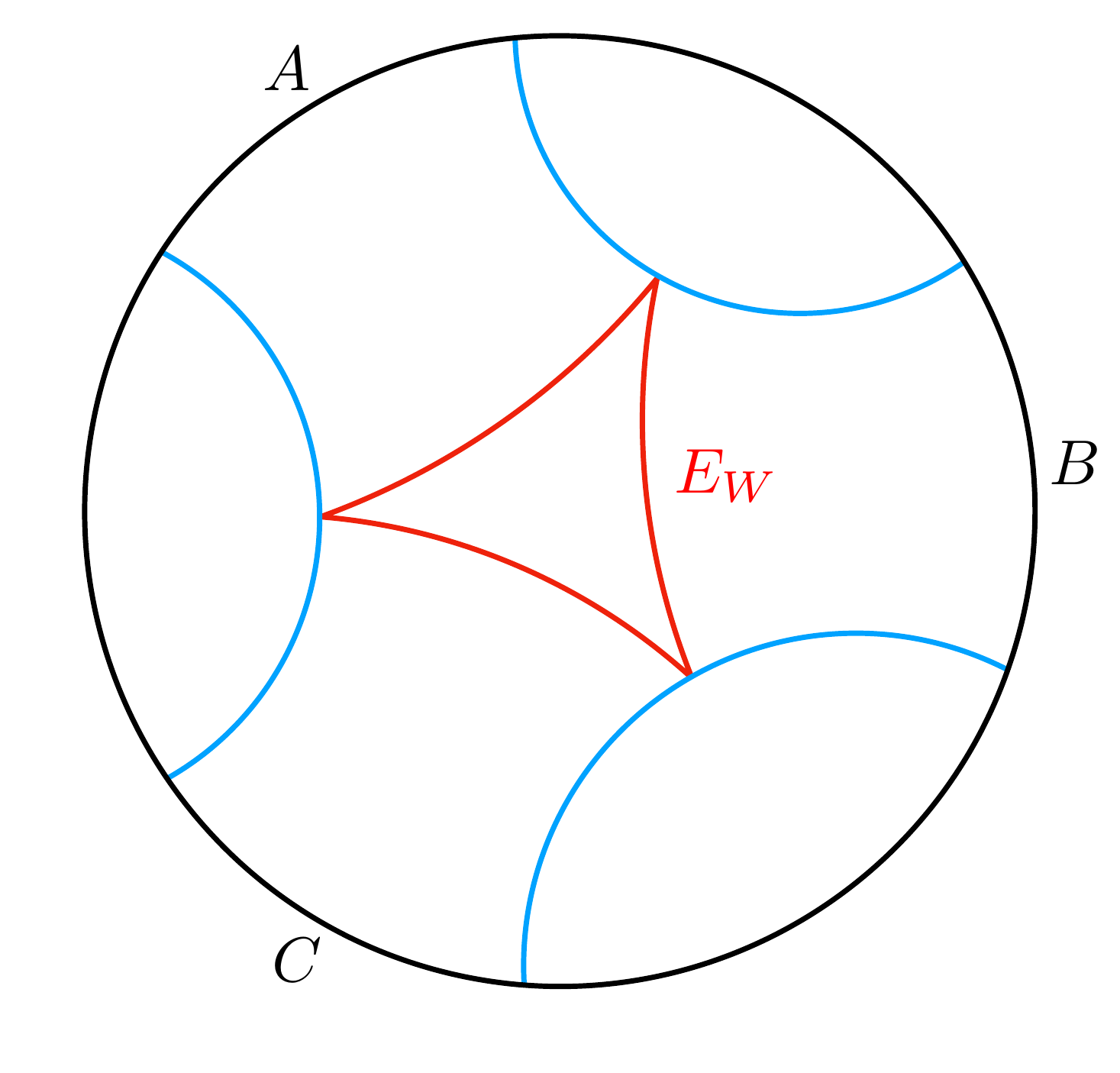}
\caption{The tripartite entanglement wedge cross section (red line) of the $ABC$ boundary subsystem, with area $E_W$. Note that the minimization here is a constrained minimization such that the red line must separate $A$, $B$, and $C$ from their respective complements in the entanglement wedge and be connected.}
\label{fig:3pep}
\end{figure}

In the tripartite case, it was conjectured in \Refs{Bao:2018gck,Umemoto:2018jpc} that the bulk object dual to this entanglement of purification is the total area of three surfaces, each of which partitions the entanglement wedge of $\rho_{ABC}$ into a section adjacent to one of the three subsystems $A$, $B$, or $C$, and a section adjacent to the remaining systems.
Furthermore, these surfaces are required to meet (in general, on codimension-three surfaces) on the Ryu-Takayanagi surfaces bounding the entanglement wedge, thus constraining the optimization.
A pictorial representation of this partitioning is given in \Fig{fig:3pep}.

The multipartite generalization of this holographic object simply extends the previous case from three surfaces to $n$ surfaces, while maintaining the meeting condition for adjacent surfaces. That is, it is conjectured that the bulk object dual to the $n$-partite entanglement of purification associated with $n$ boundary subregions $A_i$ is the minimal codimension-two polytope\footnote{Throughout, we use the straightforward generalization of ``polytope,'' ``polygon,'' etc. to denote objects with facets that are not ``flat'' in a Euclidean sense, but instead are minimal with respect to the asymptotically-AdS geometry.} with $n$ facets, for which each (codimension-two) facet is homologous to exactly one of the $A_i$, with all of the (codimension-three) ridges of the polytope located on the Ryu-Takayanagi surfaces defining the entanglement wedge. The area of this polytope in Planck units gives the entanglement of purification. In the special case of $2+1$ bulk dimensions, this surface is a minimal polygon in the hyperbolic planar geometry of the asymptotically-AdS\textsubscript{3} spacetime.

\subsection{Holographic inequalities for entanglement of purification}
\label{subsec:multi_ineqs}

We can use the geometric conjecture for the bulk dual of the multipartite entanglement of purification to derive new inequalities for holographic states in AdS\textsubscript{3}/CFT\textsubscript{2}. We will first consider a couple of relatively simple cases as a warm-up and subsequently proceed to give the general set of inequalities one can derive from our setup.

In the multipartite case, one can consider a region of the boundary comprised of the union of $2n$ simply-connected subregions $A_i$. Since there is only one spatial dimension, the $A_i$ have a natural ordering (modulo cyclic permutations and sign) from their arrangement around the boundary.
Consider the $2n$-sided bulk object that computes $E_W(A_1:A_2:\ldots:A_{2n})$ with $n>2$, and compare with the two consecutive $n$-sided objects that would compute $E_W(A_1A_2:A_3A_4:\ldots:A_{2n-1}A_{2n})$ and $E_W(A_{2n}A_1:A_2A_3:\ldots:A_{2n-2}A_{2n-1})$.
Subsets of the boundaries of these two $n$-sided objects can be formed into a $4n$-sided object circumscribing them, where each of $2n$ of the vertices is on a unique boundary-anchored minimal surface between two successive $A_i$. This circumscribing object can thus itself be deformed continuously into the minimal $2n$-sided polygon that computes the $2n$-partite $E_W$.
Hence, in this restricted case we have the inequality
\begin{equation}
\begin{aligned}
E_W(A_1:A_2:\ldots:A_{2n})&& \leq&\;\;\;\,\,  E_W(A_1A_2:A_3A_4:\ldots:A_{2n-1}A_{2n}) \\
&&& +E_W(A_{2n}A_1:A_2A_3:\ldots:A_{2n-2}A_{2n-1}).
\end{aligned}\label{eq:firstexample}
\end{equation}

This observation can be significantly generalized.
Let us next consider an $mn$-sided object defining an $mn$-partite $E_W$ and $m$ consecutive $n$-sided objects, each defining an $n$-partite $E_W$ in a manner analogous to the previous example.
The geometric construction leading to \Eq{eq:firstexample} then immediately generalizes to give
\begin{equation}
E_W(A_1:A_2: \ldots :A_{mn}) \leq \sum_{i=0}^{m-1} E_W(A_{i+1}\ldots A_{i+m}:\ldots:A_{i+1+(n-1)m}\ldots A_{i+mn}),
\end{equation}
where all indices are tacitly defined modulo $mn$.

\begin{figure}[p]
\centering
\includegraphics[scale=0.45]{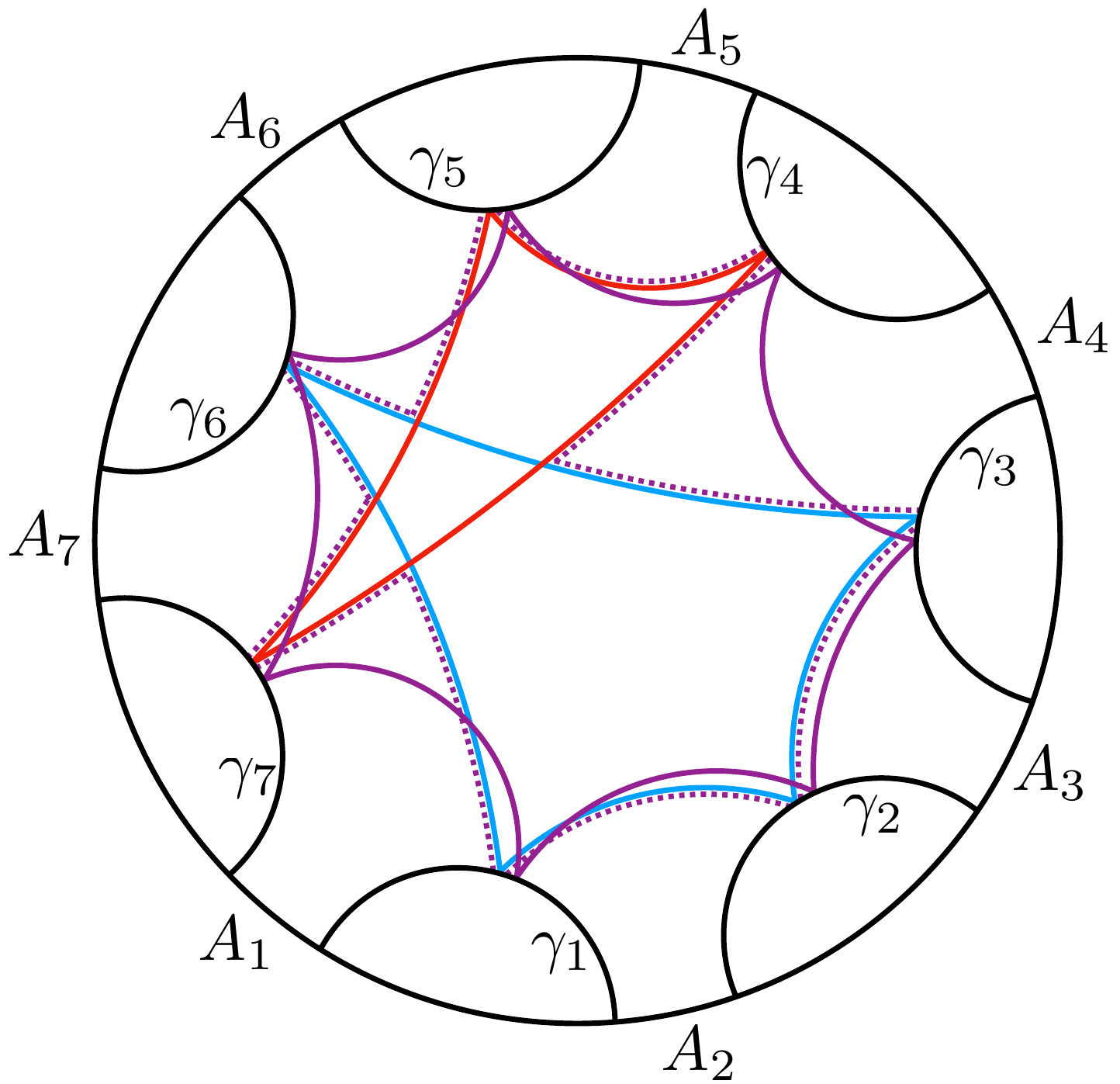}
\vspace{-5mm}
\caption{A partition of $n = 7$ boundary subregions into $n_1 = 3$ sets of subregions and $n_2 = 4$ sets of subregions. The surface $P_1$, which computes the tripartite entanglement of purification for the set $C_1 = \{A_1 A_2 A_3 A_4, A_5, A_6 A_7\}$, is shown in red, and the surface $P_2$, which computes the quadripartite entanglement of purification for the set $C_2 = \{A_2, A_3, A_4 A_5 A_6, A_7 A_1\}$, is shown in blue. Segments of these two surfaces form the polygon $\bar P$, shown with the dashed purple line, which can be deformed to the 7-partite entanglement wedge cross section (shown in solid purple).}
\label{fig:Construction1}
\end{figure}

More generally, consider an $n$-sided object whose length calculates $E_W(A_1:A_2:\ldots:A_n)$, where as before each of the regions $A_i$ is simply connected. As above, let us label the regions $A_i$ cyclically around the boundary and write the boundary-anchored minimal surface between $A_{i}$ and $A_{i+1}$ as $\gamma_i$, where all indices are defined modulo $n$.  Let us take an arbitrary partition of $n$ into $N$ parts, indexed by $j$ with sizes $n_j$, so $\sum_{j=1}^N n_j = n$, where we require each $n_j \geq 2$. We can then partition the collection of surfaces $\{\gamma_i \}$ into sets $B_j$ where $|B_j| = n_j$ and such that no union of $B_j$ is simply a collection of successive $\gamma_i$, i.e., 
\be 
\bigcup_{j\in S} B_j \neq \{\gamma_i, \gamma_{i+1},\ldots , \gamma_{i+\sum_{j\in S} n_j} \}\label{eq:Bcondition}
\ee 
for any $i,j$ and any proper subset $S$ of $\{1,\ldots,N\}$. For each $j$, define $s_j$ to be the ordered set of indices $i$ for which $\gamma_i \in B_j$, and write $s_j^k$ for the $k$th element of $s_j$. Since all indices are defined modulo $n$, for a given $j$ the $s_j^k$ are defined up to cyclic permutations of the $k$ index. We then define 
\be
C_j^k = \bigcup_{\ell = s_j^k + 1}^{s_j^{k+1}} A_\ell
\ee
for each $j,k$, $1\leq j\leq N$, $1\leq k\leq n_j$. For each $j$, drawing the minimal $n_j$-sided polygon $P_j$ connecting the $\gamma_i$ surfaces in $B_j$ corresponds to $E_W(C_j^1:C_j^2:\ldots:C_j^{n_j})$ if $n_j\neq 2$. If $n_j = 2$, then we have a $2$-gon connecting two of the $\gamma_i$, corresponding to $2E_W(C_j^1:C_j^2)$. (A minimal $2$-gon is just twice the minimal geodesic between two points.) By the condition \eqref{eq:Bcondition}, the $P_j$ are guaranteed to overlap with each other so that $\cup_j P_j$ is connected. There thus exists a polygon $\bar P$, formed by subsets of the $P_j$, that circumscribes $\cup_j P_j$, where there exists some subset of $n$ vertices of which each is on a unique $\gamma_i$, $1\leq i\leq n$. Hence, $\bar P$ can be deformed continuously into the minimal $n$-sided polygon $P$ connecting the $\gamma_i$, which computes the $n$-partite $E_W$; see \Fig{fig:Construction1}. Let us identify $j_0$ such that $n_j = 2$ for $j\in\{1,\ldots,j_0-1\}$ and $n_j >2$ for $j\in\{j_0,\ldots,N\}$. We thus have a large set of entanglement wedge inequalities,
\be
E_W(A_1:A_2:\ldots:A_n) \leq 2 \sum_{j=1}^{j_0-1} E_W(C_j^1:C_j^2) + \sum_{j=j_0}^N E_W(C_j^1:C_j^2:\ldots : C_j^{n_j}).
\ee

An example is useful. Let us consider $7$ boundary subregions. Partitioning $7$ as $3+4$, one of the identities we derive is
\be 
\begin{aligned}
E_W(A_1:A_2:\ldots:A_7) &&\leq&\;\;\;\,\, E_W(A_1 A_2 A_3 A_4:A_5:A_6 A_7) \\&&& + E_W(A_7 A_1:A_2:A_3:A_4 A_5 A_6)  .
\end{aligned}
\ee
If we let $C_j = \{C_j^k, k = 1,\dots,n_j\}$, this corresponds to taking $C_1 = \{A_1 A_2 A_3 A_4, A_5, A_6 A_7\}$ and $C_2 = \{A_2, A_3, A_4 A_5 A_6, A_7 A_1\}$.
This is illustrated in \Fig{fig:Construction1}.

If instead we take $C_1 = \{ A_2 A_3, A_4 A_5, A_6 A_7 A_1\}$ and $C_2 = \{A_1 A_2, A_3 A_4, A_5 A_6, A_7\}$, we find
\begin{equation}
\begin{aligned}
E_W(A_1 : A_2 : \cdots : A_7) &&\leq&\;\;\;\,\, E_W(A_2 A_3: A_4 A_5: A_6 A_7 A_1) \\ &&& + E_W(A_3 A_4 : A_5 A_6 : A_7 : A_1 A_2) \, .
\end{aligned}
\end{equation}
This corresponds to a different partition of $7 = 3+4$.

\begin{figure}[p]
\centering
\includegraphics[scale=0.45]{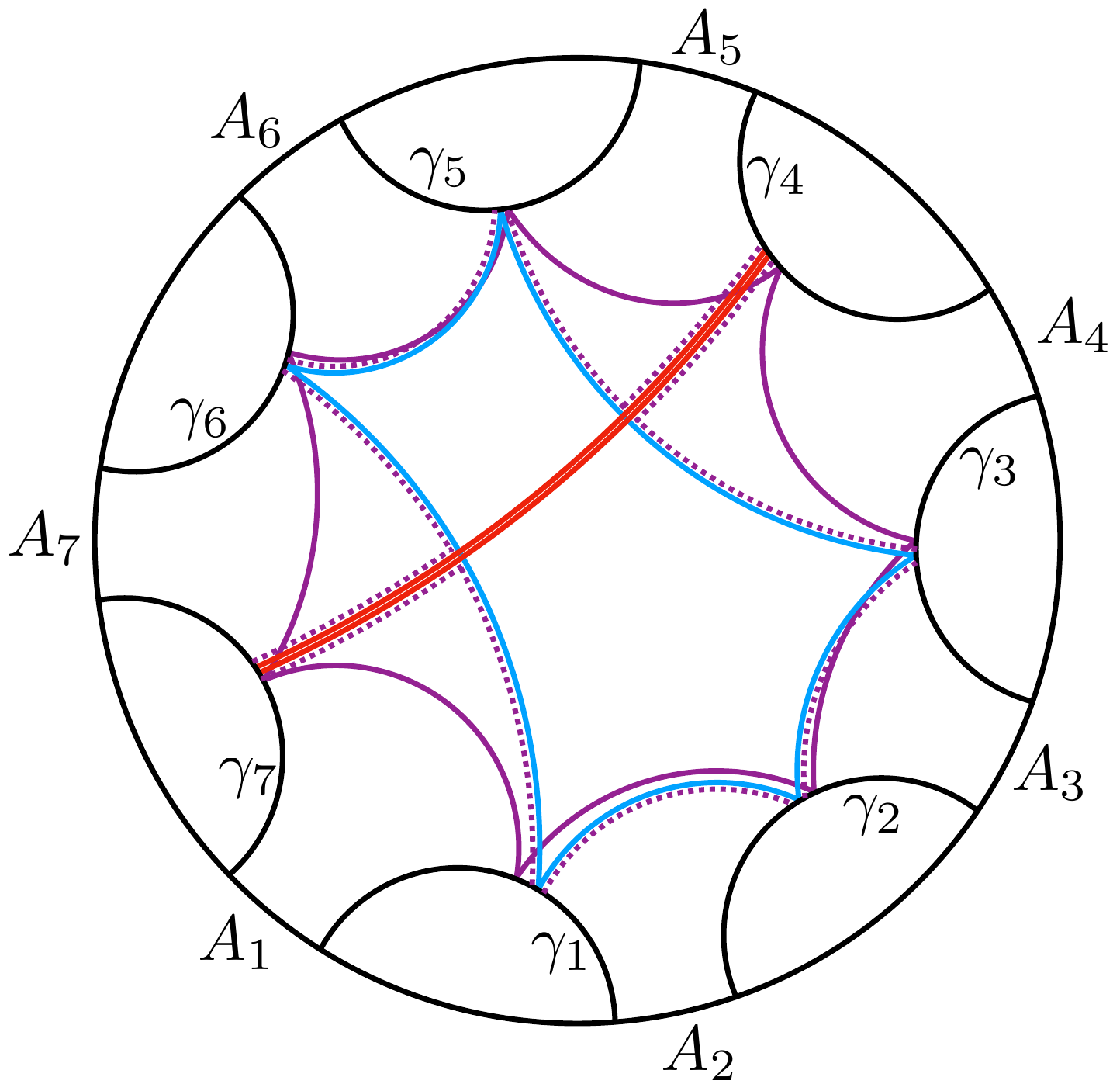}
\caption{A partition of $n = 7$ boundary subregions into $n_1 = 2$ sets of subregions and $n_2 = 5$ sets of subregions. $P_1$ (now a minimal 2-gon) is shown in red and $P_2$ is shown in blue. As before, $\bar P$ is shown in dashed purple, and the surface that computes $E_W(A_1:\ldots:A_7)$ is shown in solid purple.}
\label{fig:Construction2}
\end{figure}

\begin{figure}[ht]
\centering
\includegraphics[scale=0.45]{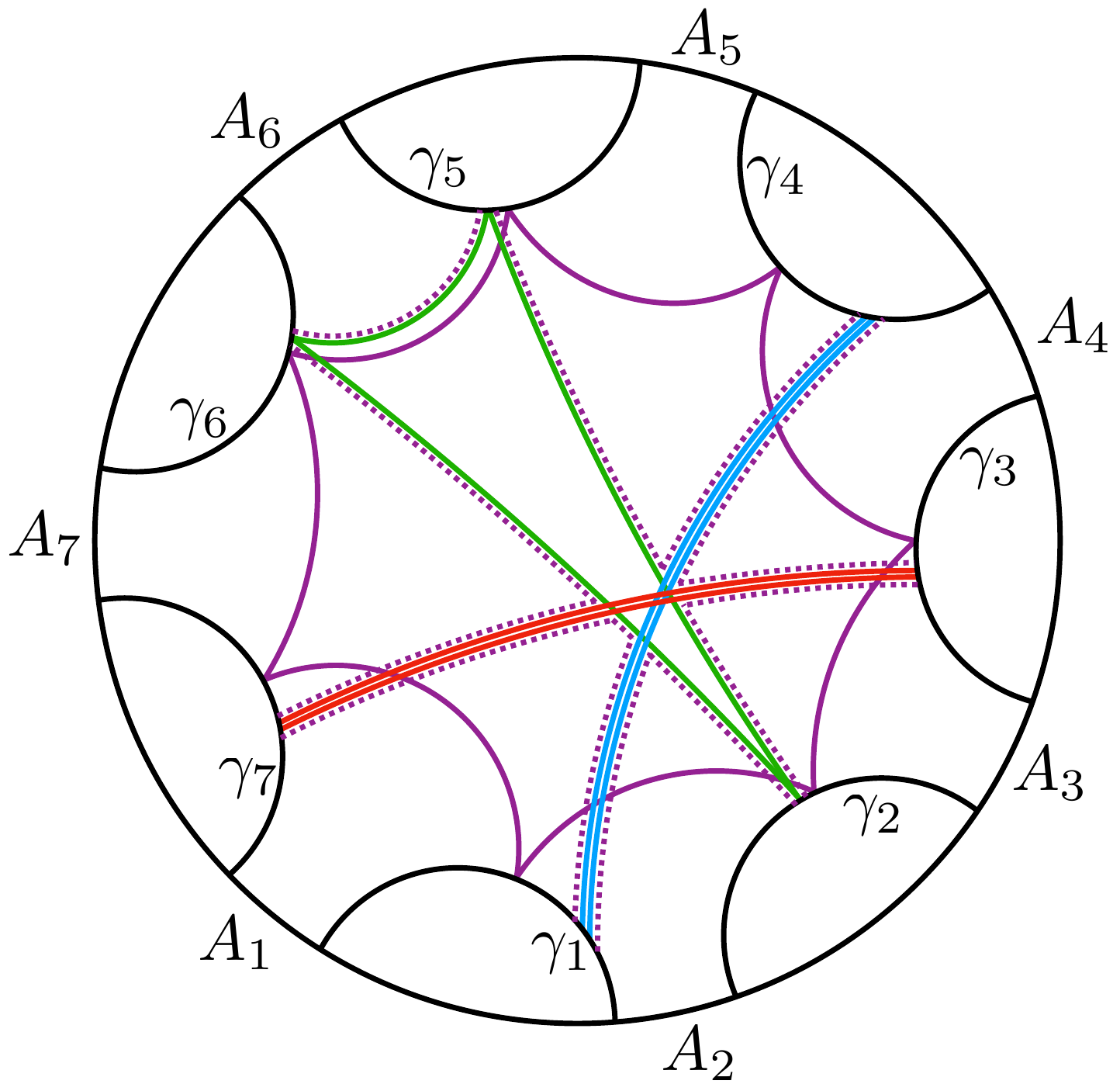}
\caption{A partition of $n = 7$ boundary subregions into $n_1 = 2$ sets of subregions, $n_2 = 2$ sets of subregions, and $n_3 = 3$ sets of subregions.}
\label{fig:Construction3}
\end{figure}

As further examples, partitioning $7$ as $2+5$, we obtain
\be 
\begin{aligned}
E_W(A_1:A_2:\ldots:A_7) &&\leq&\;\;\;\,\, 2 E_W(A_1 A_2 A_3 A_4:A_5 A_6 A_7) \\ &&& + E_W (A_7 A_1:A_2:A_3:A_4 A_5:A_6),
\end{aligned}
\ee
and partitioning $7$ as $2+2+3$, we find
\be
\begin{aligned}
 E_W(A_1:A_2:\ldots:A_7) &&\leq&\;\;\;\,\,2E_W (A_1 A_2 A_3:A_4 A_5 A_6 A_7) \\
 &&&+ 2E_W(A_2 A_3 A_4:A_5 A_6 A_7 A_1) \\
 &&&+ E_W (A_7 A_1 A_2:A_3 A_4 A_5:A_6),
 \end{aligned}
\ee
along with many other possible choices of the assignment of the $A_i$ into subsets.
These last two partitions are illustrated in Figs.~\ref{fig:Construction2} and \ref{fig:Construction3}, respectively.

\section{Constructing Wormhole Geometries via Identifications of AdS\textsubscript{3}}
\label{sec:WHviaOrbifolds}

\begin{figure}[h]
\centering
\includegraphics[scale=0.45]{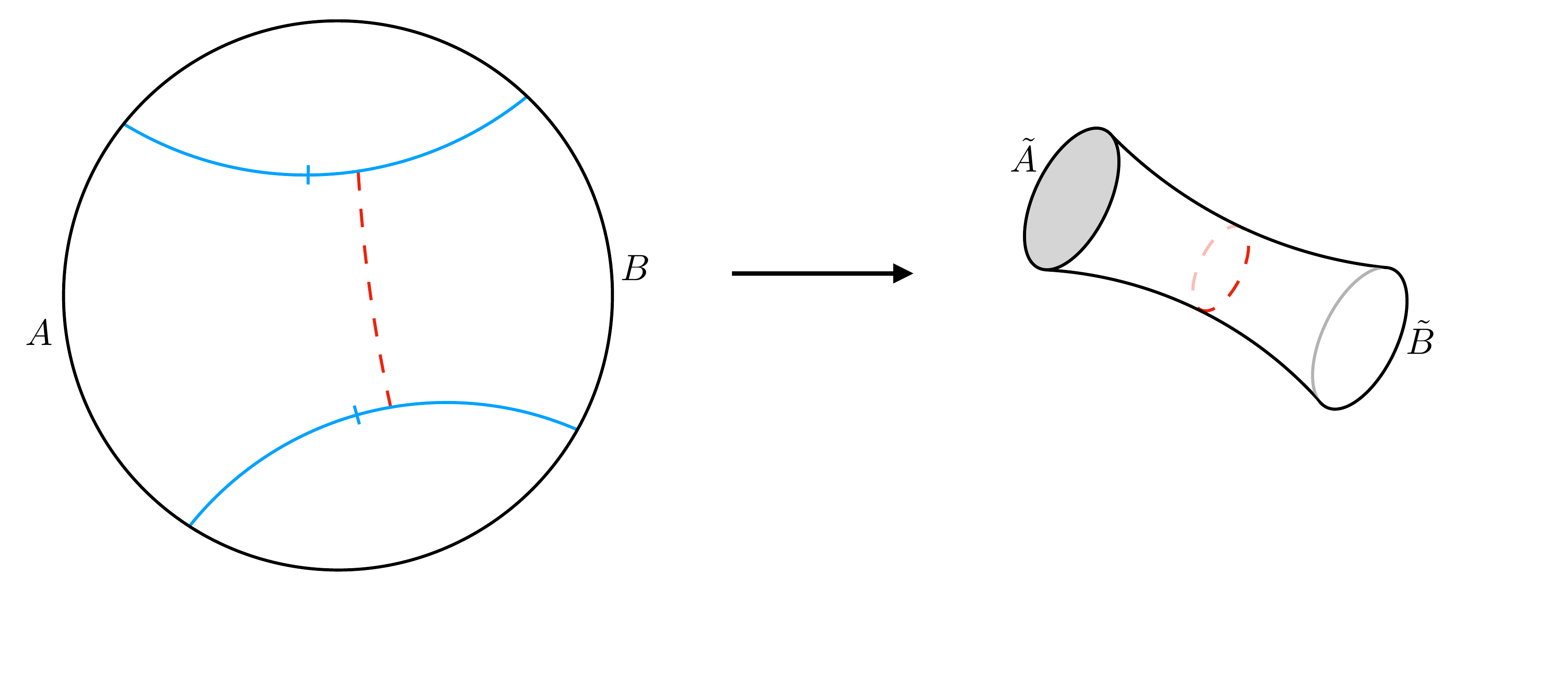}
\vspace{-1cm}
\caption{The two-surface identification used to make the two-sided wormhole geometry. Notches indicate the identification of the two blue surfaces with each other in the left subfigure, creating the wormhole geometry depicted in the right subfigure. Crucially, the minimal surface connecting the identified geodesics (depicted here by a red dashed line) maps onto the horizon area in the wormhole geometry.}
\label{fig:2sorb}
\end{figure}

In this section, we briefly review how multi-black-hole spacetimes are obtained by taking a quotient of pure AdS\textsubscript{3} by a set of isometries.
Starting with the Poincar\'e disk,\footnote{Any totally geodesic, spacelike slice will do, but without loss of generality, we can work entirely with the Poincar\'e disk since the former always maps onto the latter by stereographic projection. A submanifold is totally geodesic if any geodesic on the submanifold (with respect to the induced metric) is also a geodesic in the parent manifold.} taking a quotient by a single hyperbolic isometry produces a new Riemannian manifold that everywhere has constant negative curvature.
The resulting manifold is the same as if one had cut a strip bounded by boundary-anchored geodesics out of the disk and glued it shut, as illustrated in \Fig{fig:2sorb}. 
This new spacelike manifold, which consists of a wormhole that connects two asymptotically-hyperbolic regions, then serves as the initial condition for a new locally asymptotically-AdS spacetime.
Quotienting by more than one isometry produces more complicated multiboundary wormholes. 
Here, we only sketch the process with a small number of mathematical details, but a complete treatment of the topic may be found in \Refs{Brill:1995jv,Aminneborg:1997pz,Brill:1998pr}.
In AdS/CFT, the quotient process produces a corresponding CFT state.
We will keep the discussion of the CFT-related aspects minimal here (restricting ourselves to what can be gleaned from the Ryu-Takayanagi formula), but these are elaborated in great detail in \Refs{Balasubramanian:2014hda, Marolf:2015vma}.

First, consider two boundary-anchored geodesics in the Poincar\'e disk that do not share any endpoints, as shown in \Fig{fig:2sorb}.
Then there exists a unique isometry of the disk that bijectively maps points on the first geodesic onto their closest points on the second.
One can show that this isometry has no fixed points in the strip between the two geodesics nor on the portion of the boundary between the two geodesics~\cite{Brill:1998pr}.
Moreover, the image of this strip under repeated applications of the isometry or its inverse covers the whole Poincar\'e disk, accumulating only at two fixed points on the boundary that lie in the regions subtended by the boundary-anchored geodesics.
Therefore, if we take a quotient of the disk and identify points that are related by the action of this isometry and its inverse, the result is a smooth, nonsingular manifold.

Equivalently, one may think of the quotient as taking the initial strip, or \emph{fundamental domain}, and periodically identifying its two geodesic boundaries.
Following the identification, the original boundary subregions between the two geodesics are mapped onto two complete AdS boundaries that are joined by a smooth wormhole.
The throat of this wormhole is a segment of the unique geodesic that is left invariant by the isometry in the original Poincar\'e disk.
(The fixed points of the isometry are therefore the points where this invariant geodesic meets the conformal boundary.)
Therefore, instead of specifying a pair of geodesics to be identified, one can equivalently specify the unique invariant geodesic and the size of the desired wormhole throat to characterize a quotient of the Poincar\'e disk.

In principle, one is not limited to identifying a pair of geodesics; one could also begin with a smooth curve that does not intersect its image under the isometry.
Geodesics are merely a convenient choice of fundamental domain boundary that guarantees the smoothness of the resulting metric.
They are furthermore useful for constructing more complicated wormholes.

\begin{figure}[ht]
\centering
\includegraphics[scale=0.45]{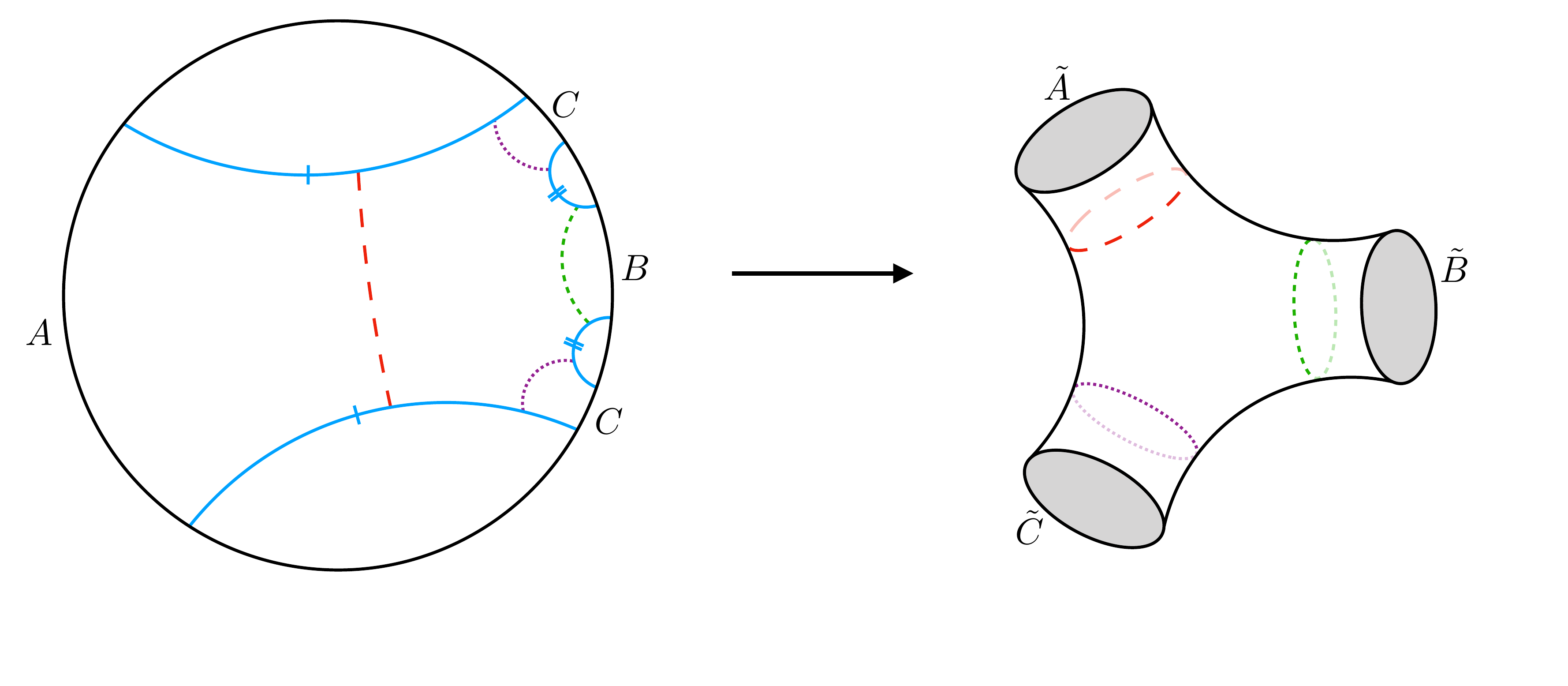}
\vspace{-1cm}
\caption{The two different pairwise surface identifications used to make the three-sided wormhole geometry. Single or double notches indicate the identifications of pairs of blue surfaces in the left subfigure, creating the wormhole geometry depicted in the right subfigure. The minimal surfaces that connect pairs of identified geodesics (depicted here by dashed lines of differing colors and weights) map onto the respective horizon areas (shown with the same dashed lines) in the wormhole geometry.}
\label{fig:3sorb}
\end{figure}

The construction described above produces a time-symmetric slice of a two-sided BTZ black hole \cite{Banados:1992wn}, which one could call the $t=0$ slice in the usual AdS-Schwarzschild coordinates.
By iterating the process---that is, by taking a quotient with respect to several isometries---one obtains more complicated geometries that consist of several asymptotic regions that are joined by a common wormhole, as illustrated in \Fig{fig:3sorb}.
In the equivalent cutting-and-gluing picture, the process consists of specifying a fundamental domain using an even number of boundary-anchored geodesics and then identifying pairs of these geodesics.

\begin{figure}
\centering
\includegraphics[scale=0.45]{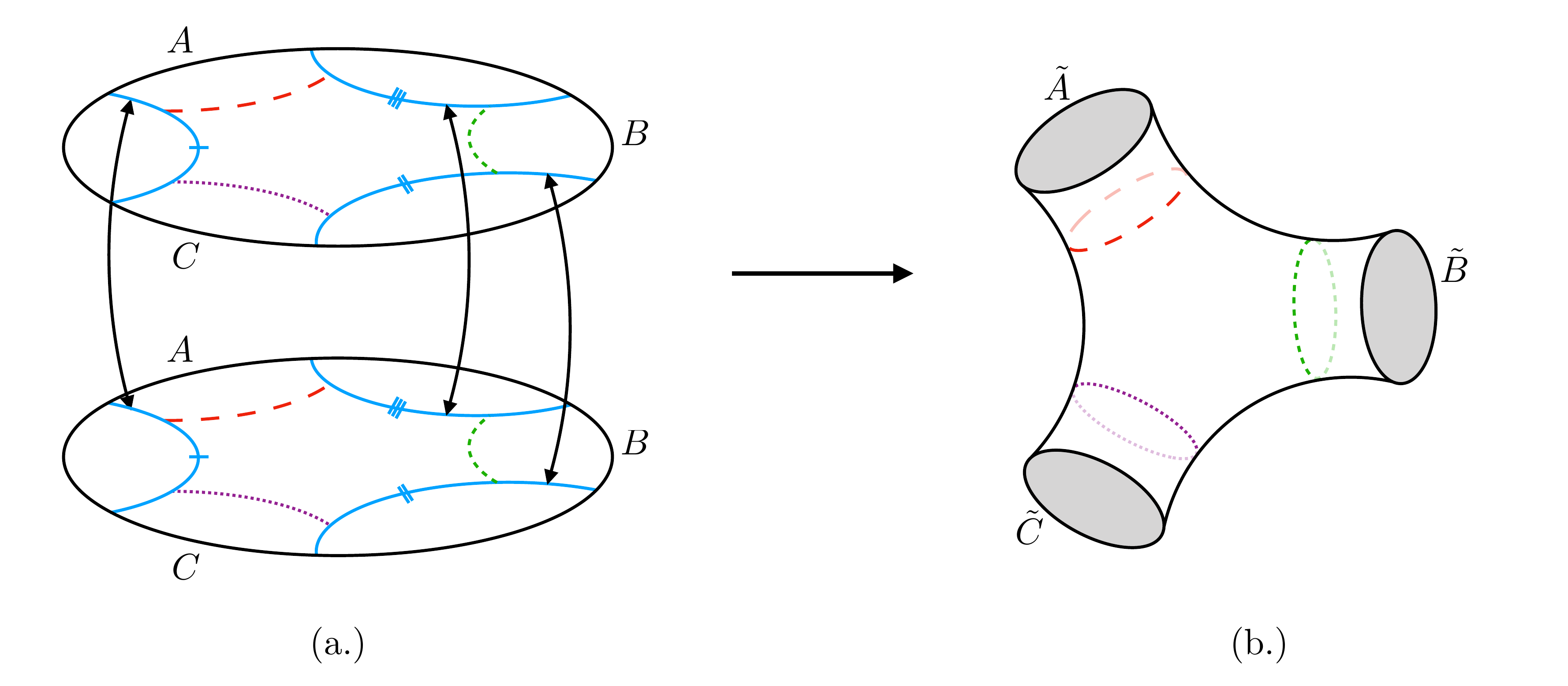}
\vspace{-7mm}
\caption{The doubling procedure for generating a multiboundary wormhole, illustrated here for $k=3$ boundary subregions. (a) A single Poincar\'e disk containing the original subregions is first doubled. Identifying boundary-anchored geodesics (blue lines) between the original and the double produces a 3-sided wormhole (b). The horizons in the wormhole throats (dashed lines) correspond to geodesic segments in the original pre-quotient geometry.}
\label{fig:doubling}
\end{figure}

\emph{Doubling} \cite{Brill:1995jv,Brill:1998pr} is an algorithmic way to construct a multiboundary wormhole geometry with a prescribed set of properties, such as number of asymptotic regions, throat sizes, and distances between throats.
To illustrate this process, consider $k$ boundary subregions on the Poincar\'e disk and $k$ adjacent boundary-anchored geodesics that connect the endpoints of adjacent boundary subregions.
This is illustrated for $k=3$ in \Fig{fig:doubling}.
Next, make a copy of the disk and the geodesics.
By cutting along the geodesics and gluing the original cut-out to its copy, one obtains a wormhole geometry with $k$ mouths.
This geometry is a time-symmetric initial condition for $k$ asymptotically AdS regions, each of which contains a black hole.
The black holes are connected by a wormhole with $k$ mouths whose structure is hidden behind the black hole horizons.

While it may not be immediately apparent from this cutting-and-gluing picture, doubling still consists of an isometric quotient.
This can be seen as follows.
In the original disk, pick one of the boundary-anchored geodesics.
There exists an isometry that moves this geodesic onto the diameter of the Poincar\'e disk.
After applying this isometry, the copy of the original region can be embedded into this single Poincar\'e disk as the reflection of the (isometrically deformed) region across the diameter of the disk.
This realizes the original/copy identification from the doubling picture for the first geodesic that we selected.
The other original/copy pairs of geodesics are now related by isometries of the single Poincar\'e disk, and so doubling is indeed a quotient by isometries.
See in particular Fig.~3b in \Ref{Brill:1998pr} for illustration.

As before, the throats of the resulting wormhole correspond to unique geodesic segments that intersect the adjacent geodesics in the pre-identification geometry.
The size of a wormhole throat is therefore equal to twice the length of the corresponding geodesic segment in a single copy of the pre-identification geometry.
Similarly to the two-boundary case, a $k$-boundary wormhole can be specified by choosing these geodesic segments; their lengths and relative separations correspond to specifying the black hole masses and the separations of their horizons in the wormhole.
These $2k$ parameters are not all independent; however it is possible to independently specify all $k$ black hole masses for $k \geq 3$.
Further details are explained in \Ref{Brill:1998pr}.

Within AdS/CFT, for a wormhole with $k$ boundaries, the quotient construction yields a corresponding state on the tensor product of $k$ CFTs that one can identify with the $k$ asymptotic boundaries. 
If one were to trace out all but one of the CFT boundaries, the result is a thermal state that is dual to a single-sided BTZ black hole of temperature given by the horizon size.
In the full wormhole geometry, the horizons are the minimal surfaces that are homologous to each entire boundary and thus compute the entanglement entropies of the reduced states on each individual boundary in the CFT.

\section{Interpreting Entanglement of Purification via Wormholes}
\label{sec:interpretingEP}

It is now clear that in the context of three-dimensional holographic theories, the bulk object whose area is conjectured to be dual to entanglement of purification and the minimal geodesic segments that define the isometric identifications are very closely related.
When the entanglement wedge connects two boundary subregions they are precisely the same, and $E_W$ in the pre-identification geometry is equal to the entanglement entropy of an entire CFT boundary in the wormhole geometry.
When $E_W$ is the sum of the areas of a union of disjoint simply-connected bulk geodesics, then it is given precisely by the sum of entanglement entropies of a collection of CFT boundaries in the wormhole state, which in turn are computed by these minimal surfaces.

The upshot of this observation is quite interesting. At least for geometries that have enough isometries to allow for quotienting, such as empty AdS\textsubscript{3}, the way to compute entanglement of purification in $(1+1)$-dimensional holographic CFTs appears to be via calculation of entanglement entropies of the CFTs constructed by isometric identification.
Calculations of these entanglement entropy quantities is much more straightforward than direct calculation of the entanglement of purification, as in the work of \Ref{Calabrese:2004eu}.
Perhaps such a statement (or a variant thereof) can even be proven using CFT alone, without the need to appeal to an explicit holographic construction.

Assuming the $E_P=E_W$ conjecture---and our multipartite generalization---is correct, one can reach several conclusions about $E_P$ for such states, as we will now see. 

\subsection{Entanglement of purification inequalities from holographic entanglement entropy}

All entanglement entropy inequalities that hold for any holographic state by definition hold in holographic wormhole geometries.
In particular, in a wormhole geometry, we are free to consider the entanglement entropy of entire connected boundaries.
The minimal Ryu-Takayanagi surfaces that compute the entanglement entropy of entire boundaries, however, now have an additional interpretation as entanglements of purification in the pre-identification geometry.
Therefore, inequalities that constrain entanglement entropies in the wormhole geometry yield corresponding constraints on the entanglement of purification.
Consider, for example, subadditivity:
\begin{equation}
S(\tilde{A})+S(\tilde{B})\geq S(\tilde{A}\tilde{B}).
\end{equation}
Here, we use letters with tildes for the post-identification CFT boundaries and letters without tildes for the pre-identification boundary subregions.
Let us consider the simplest nontrivial case, that of a three-sided wormhole with boundaries $\tilde{A}, \tilde{B}$, and $\tilde{C}$.
By substitution, we have immediately that
\begin{equation}
E_W(A:BC)+E_W(B:AC)\geq E_W(C:AB).
\end{equation}
Note that this does not follow from known properties of $E_P$, as in this case $\rho_{ABC}$ is not pure (since it is a state corresponding to a proper subset of AdS\textsubscript{3}).
In a similar fashion, further holographic entanglement entropy inequalities, such as strong subadditivity or the Araki-Lieb inequality, convert into a constraint on $E_P$ in the pre-identification CFT(s). 

\begin{figure}[ht]
\centering
\includegraphics[scale=0.45]{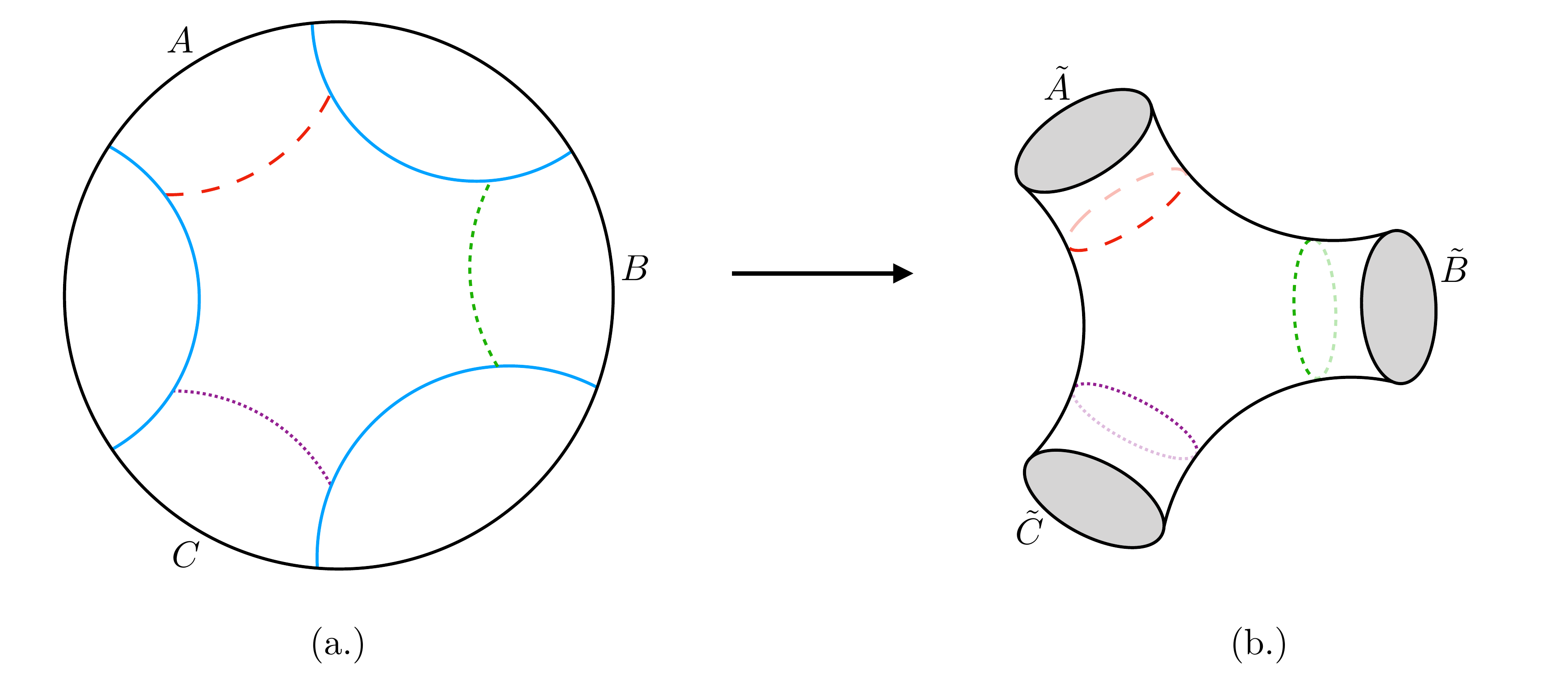}
\vspace{-4mm}
\caption{The completely connected phase, where the entanglement wedge of $ABC$ is a single connected region bounded by the blue Ryu-Takayanagi surfaces. The bipartite entanglement wedge cross sections (dashed lines) map onto wormhole throats in the quotient geometry.}
\label{fig:3party-3}
\end{figure}

\begin{figure}[p]
\centering
\includegraphics[scale=0.45]{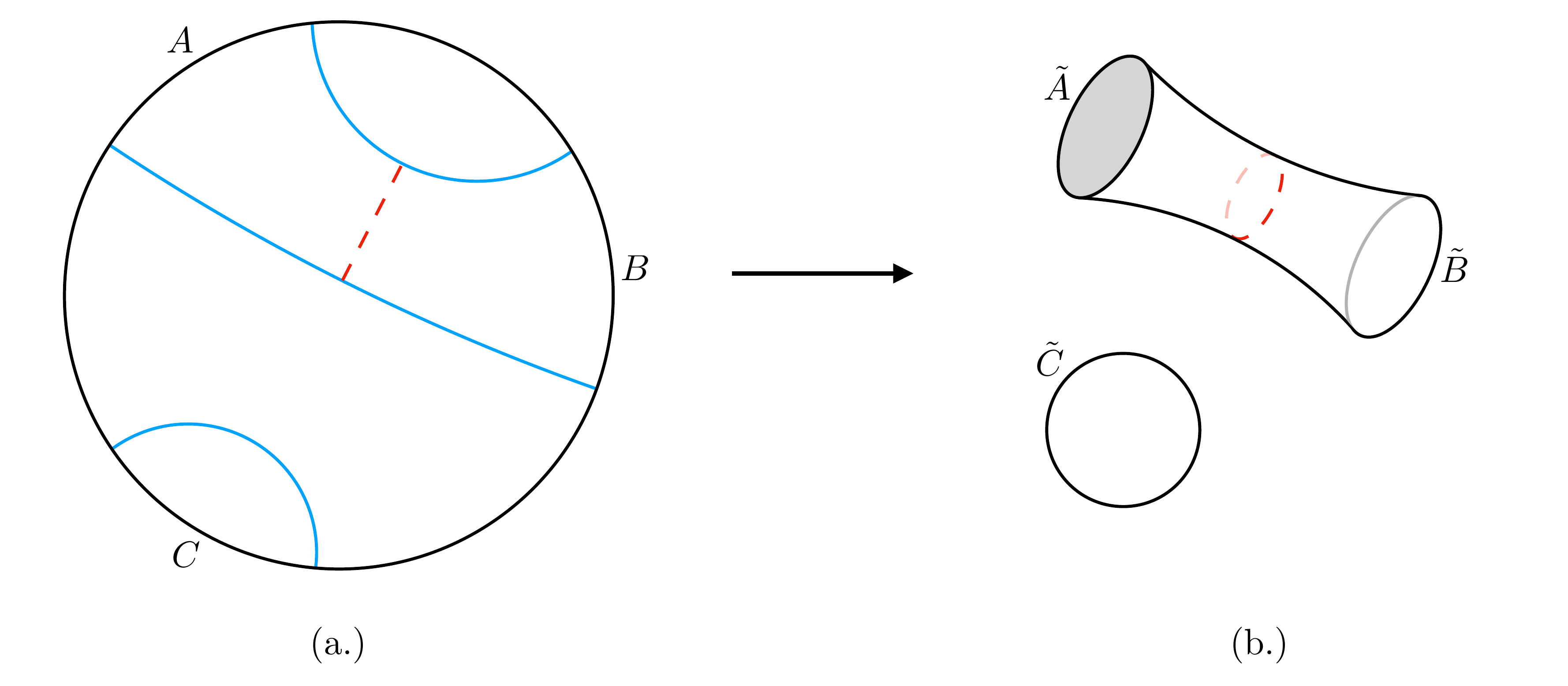}
\vspace{-4mm}
\caption{Another phase, in which the entanglement wedge of $ABC$ is a disjoint union of the entanglement wedge of $AB$ and that of $C$. The entanglement wedge for $AB$ is mapped onto a two-sided wormhole and its entanglement wedge cross section (red dashed line) is mapped to the Ryu-Takayanagi surface associated with a complete boundary, while $C$ is mapped to a trivial hyperbolic geometry.}
\label{fig:3party-2}
\end{figure}

\begin{figure}[p]
\centering
\includegraphics[scale=0.45]{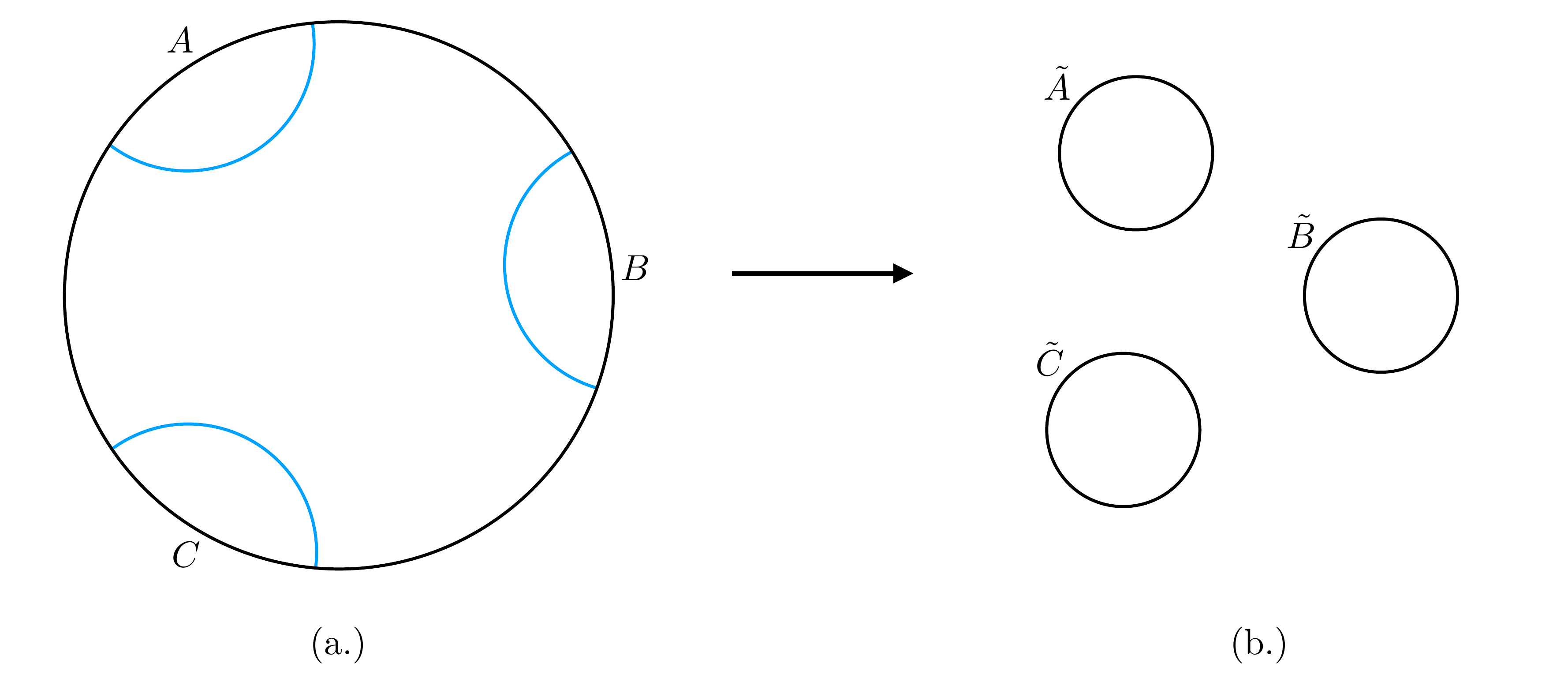}
\vspace{-4mm}
\caption{The totally-disconnected phase, when the entanglement wedge of $ABC$ is the disjoint union of the entanglement wedges of $A$, $B$, and $C$ individually. Each boundary subregion is mapped to a disconnected trivial geometry.}
\label{fig:3party-1}
\end{figure}

In order to convert entanglement entropy inequalities into inequalities for the entanglement of purification, one must show that any $E_W$ surface can be mapped onto a Ryu-Takayanagi surface in a wormhole geometry via isometric identification.
To begin, consider the three-party case in which we choose three simply-connected boundary subregions in the parent geometry.
The different possible topologies for the entanglement wedge, as well as the entanglement wedge cross sections, are drawn in Figs.~\ref{fig:3party-3}a--\ref{fig:3party-1}a.

In the first configuration (\Fig{fig:3party-3}a), all of the bipartite $E_W$s formed from $A$, $B$, and $C$ are nonzero.
In this case, each entanglement wedge cross section can be mapped onto (one half of) a wormhole throat in a 3-sided wormhole via the doubling procedure described in the previous section.
As such, any bipartite $E_W$ is converted into the length of one of the wormhole throat openings, which is the Ryu-Takayanagi surface for one of the full boundaries.

In the second configuration (\Fig{fig:3party-2}a), the entanglement wedge decomposes into two disconnected components.
Here, $E_W(A:B) = E_W(A:BC) = E_W(B:AC) \neq 0$ and $E_W(C:A) = E_W(C:B) = E_W(C:AB) = 0$.
If we identify the geodesics that are anchored to the boundaries of $A$ and $B$, then the nonzero $E_W$s are mapped onto the length of the throat of the wormhole that joins $\tilde A$ and $\tilde B$ in the resulting two-sided wormhole geometry.
Since $E_W$ of $C$ with anything else is zero, we should independently map the state on $C$ onto a pure state on the conformal boundary $\tilde C$ of a disconnected space.
Then, we can think of the trivial $E_W$s as being calculated by the trivial Ryu-Takayanagi surface that subtends all of $\tilde C$.
The simplest map is to just let the state on $\tilde C$ be a copy of the state on the complete original boundary, but in general, the state on $\tilde C$ could be (in principle, a unitary rotation of) the joint state of $\rho_C$ and a purification.

The last possibility is the trivial configuration in which the entanglement wedge consists of three disconnected regions.
Here, every $E_W$ vanishes.
In this case, we can think of the $E_W$s as being calculated by (trivial) Ryu-Takayanagi surfaces for three disconnected spacetimes as shown in \Fig{fig:3party-1}b (one for each initial boundary subregion).

~

The case of an arbitrary number of boundary subregions follows in essentially the same way.
Consider a holographic CFT with $k$ boundary subregions $A_1, \dots, A_k$.
Partition these boundary subregions into sets $X_a,$ $1 \leq a \leq n$, such that the entanglement wedges of $\cup_{A_i \in X_a} A_i$ are the connected components of the full entanglement wedge of $\cup_{j = 1}^k A_j$.
For each $X_a$, form a disjoint wormhole geometry by doubling (or a trivial disconnected geometry when $|X_a| = 1$) with a full boundary $\tilde A_i$ for each $A_i \in X_a$.
Then it follows that any bipartite $E_W$ made up of any of the $A_i$s is given by the entanglement entropy of the corresponding $\tilde A_i$s.

It is also clear that the geometric quantity conjectured to be dual to tripartite (and by analogy multipartite) $E_P$ also has a nontrivial lower bound.
In particular, note that in the quotient geometry, each of the three minimal surfaces that partition the pre-identification entanglement wedge transform into non-minimal surfaces homologous to the individual boundary regions, and thus are each individually larger than the minimal surfaces that compute bipartite $E_P$s.
This geometric observation implies
\be 
E_P(A:B:C)\geq E_P(A:BC)+E_P(B:AC)+E_P(C:AB),
\ee
reproducing Proposition 12 of \Ref{Umemoto:2018jpc}.

\subsection{Holographic entanglement entropy inequalities from entanglement of purification}

One can also obtain new results by running our construction in the other direction,  to convert inequalities for the entanglement of purification into entanglement entropy inequalities.
For example, recall that $E_P(A:BC)\geq E_P(A:B)$ for all quantum states and that \linebreak $E_P(AC:BD)\geq E_P(A:B)+E_P(C:D)$ for states with holographic duals, assuming the $E_P=E_W$ conjecture \cite{Takayanagi:2017knl,Nguyen:2017yqw}.
By replacing these entanglements of purification with entanglement entropies, we get from the first inequality that
\begin{equation}
S(\tilde{A})_{\tilde{A}\tilde{B}\tilde{C}}\geq S(\dbtilde A)_{\dbtilde{A}\dbtilde{B}},
\end{equation}
where the left-hand side is the entanglement entropy of $\tilde{A}$ in the three-sided wormhole $\tilde{A}\tilde{B}\tilde{C}$ and the right-hand side is the entanglement entropy between the two boundaries of the two-sided wormhole $\dbtilde{A}\dbtilde{B}$ that one forms via the appropriate identifications on the original geometry. 
Adopting similar notation, the second inequality gives that
\begin{equation}
S(\tilde A \tilde C)_{\tilde{A}\tilde{B}\tilde{C}\tilde{D}}\geq S(\dbtilde{A})_{\dbtilde{A}\dbtilde{B}}+S(\dbtilde{C})_{\dbtilde{C}\dbtilde{D}}.
\end{equation}
Inequalities of this type are not typically encountered in relativity; they relate the sizes of minimal surfaces in different geometries.
The reason why such quantities appear here is that these geometries are descended from the same parent geometry, and so these unusual inequalities correspond to standard comparisons of areas within a single pre-identification parent geometry.

Similarly to how we showed that entanglement wedge cross sections can always be mapped onto Ryu-Takayanagi surfaces in a wormhole geometry in the previous subsection, here we run the argument in the other direction and show that any multiboundary wormhole geometry can be turned into an entanglement wedge setup in a Poincar\'e disk.
This establishes that inequalities of the types shown above hold for multiboundary wormhole states in general.
Consider a wormhole with $k$ asymptotic boundaries.
This manifold is topologically equivalent to a sphere with $k$ punctures.
Arrange the punctures in a circle around the sphere's equator, and make a one-cycle graph by drawing $k$ edges on the sphere, connecting adjacent punctures.
Next, cut the sphere along the edges of the graph.
This divides the sphere into two disks, with each puncture mapped onto identified pairs of regions on the boundaries.
The cuts correspond to the pairwise-identified geodesics in the bulk of the disks.
The two disks constitute the doubled version of the single Poincar\'e disk generating the wormhole geometry.

\subsection{Optimal purification from the surface-state correspondence}

In the analysis of the entanglement of purification, a good guiding principle to keep in mind is the conjectured surface-state correspondence \cite{Miyaji:2015yva}.
The surface-state correspondence is the proposition that one can map the state on the full boundary onto a state on a closed surface in the bulk by applying isometries that correspond to continuously deforming the boundary into the closed bulk surface in question.
Because these isometries preserve the state purity\footnote{One should think of the isometries as being realized by a unitary operation $V$, on a subspace of a Hilbert space $\Hil_1 \otimes \Hil_2$, that schematically acts as $V: \ket{\psi}_{12} \mapsto \ket{\tilde \psi}_1 \otimes \ket{0}_2$.} when mapping from the full boundary to a homologous closed surface, a pure state on the complete boundary can be pushed to a pure state on the closed surface in the bulk.
If there is a black hole in the geometry, such that the boundary state is a mixed state, then the conjecture is that it is possible to push to a mixed state into the bulk.
In either case, the dimension of the Hilbert space associated with the closed bulk surface is the exponential of the surface's area in Planck units.

\begin{figure}[h]
\centering
\includegraphics[scale=0.45]{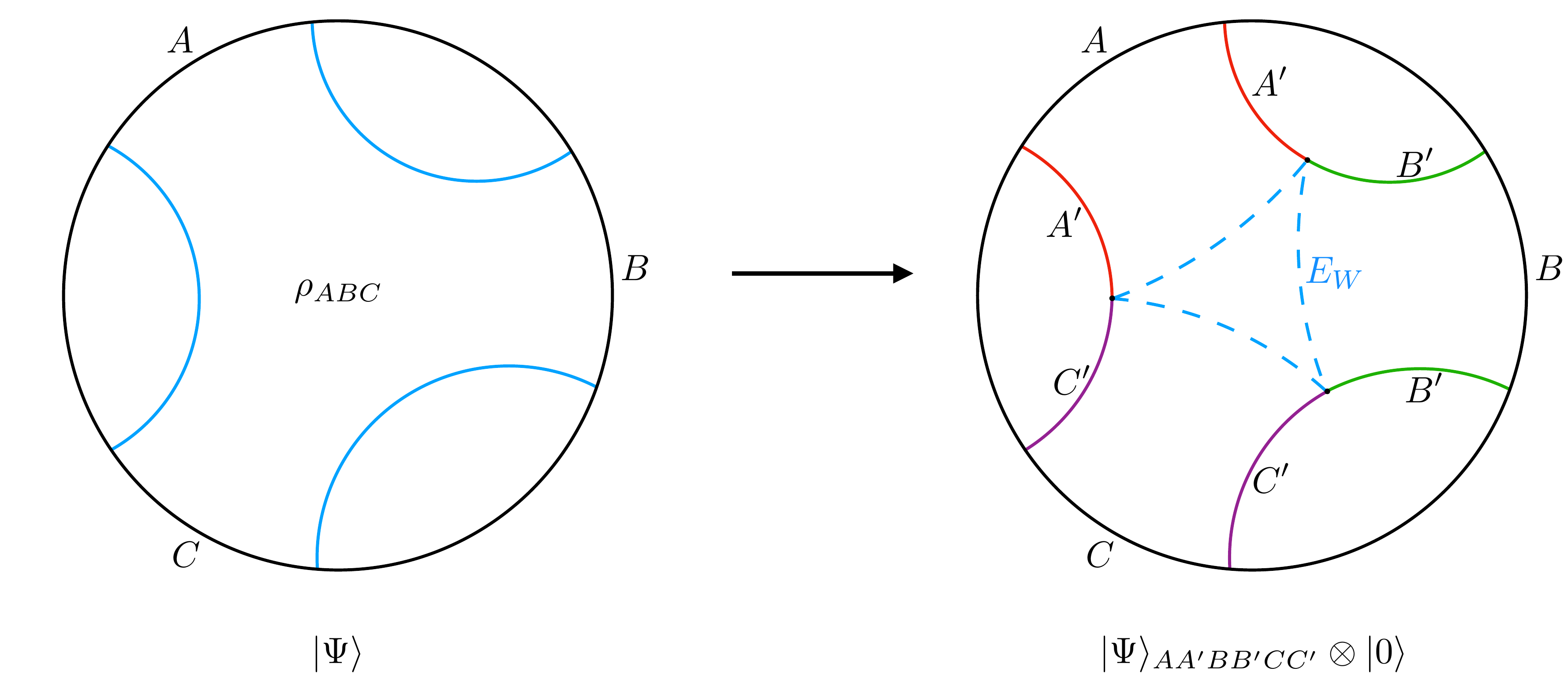}
\caption{Illustration of the pushing procedure. The entanglement wedge of $ABC$ (bounded by the blue surfaces) is described by the reduced density matrix $\rho_{ABC}$ in the left subfigure, while the entire boundary is described by the pure state $\ket{\Psi}$. Using the surface-state correspondence and $E_P = E_W$ conjecture in combination, we identify the tripartite entanglement wedge cross section (dashed blue line) and identify the purifying systems $A'$, $B'$, $C'$ as living on the union of the portions of the Ryu-Takayanagi surfaces on the $A$, $B$, or $C$ side of the entanglement wedge cross section, respectively (red, green, and purple surfaces). The full entanglement wedge is now described by $\ket{\Psi}_{AA'BB'CC'}$, while the extra boundary degrees of freedom unentangled with the entanglement wedge are described by some state $\ket{0}$.}
\label{fig:surfacestate}
\end{figure}

In the context of the $E_P=E_W$ conjecture, the surface-state correspondence produces a particular set of purifying $A_i'$ systems for which each $A_i'$ is identified with the portion of the Ryu-Takayanagi surface on the same side of the adjacent geodesics as $A_i$; see \Fig{fig:surfacestate}. 
The joint pure state defined by the unions of the $A_iA_i'$ is the state obtained via the surface-state correspondence from the entire boundary, where pushing is only done on the boundary regions that make up the complement of the union of the $A_i$s.

This idea also suggests a natural identification for the Hilbert spaces of the CFTs in the wormhole geometries created by the quotient procedure, if we make the further conjecture that the quotient does not introduce additional Hilbert space factors not associated with the boundary of the entanglement wedge.
Namely, it is natural to conjecture that the $\tilde A_i$ are isomorphic to the $A_iA_i'$ themselves.
The states that correspond to these wormhole geometries are pure, as is the state defined on the union of the $A_iA_i'$.
Note that the quotient procedure also preserves on which side of the entanglement wedge cross sections the boundary regions (and portions of RT surfaces) lie.
Because the only Hilbert space factors in the wormhole geometries are the CFT boundaries themselves, this strongly suggests that each CFT boundary corresponds to a single $A_iA_i'$. 

In the case of two boundary subregions $A$ and $B$, this proposed identification leads to the immediate conclusion that, if $E_P=E_W$ holds, the wormhole geometry calculating the entanglement of purification is the optimal purification of $\rho_{AB}$ that corresponds to a classical bulk geometry.
By optimality, we mean that it has minimal Hilbert space dimension and that it achieves the infimum in the definition of $E_P$ (cf. \Eq{eq:muliepdef}).

Consider the purification of $\rho_{AB}$ given simply by the pure state in the entire original CFT.
Because the mixed state defined on the complement of $AB$ on the boundary was not maximally entangled with $\rho_{AB}$, much of the Hilbert space is wasted by this purification of $\rho_{AB}$, even when $S(AA')$ is extremized.
Once one restricts to the state defined only on the Hilbert space factor $AA'BB'$, however, there is no more waste: because the density of degrees of freedom in $A'B'$ is given by the length in Planck units of the geodesics with which $A'B'$ is identified, i.e., $\log \dim A'B'=S(AB)$, $A'B'$ has the minimal possible Hilbert space dimension that can purify $\rho_{AB}$.
Because the $E_P=E_W$ conjecture, combined with the quotient procedure, gives us that this is both the state defined on the wormhole and that its throat calculates the entanglement of purification, we have our claim that it is a purification of minimal Hilbert space dimension.
Morevoer, in order for $E_P = E_W$ to hold, it must be that the state on $A'B'$ achieves the infimum in the definition of $E_P$, since $E_P(A:B) = S(\tilde A)$ and $S(\tilde A) = S(AA')$.
The insistence on there being a corresponding classical bulk geometry eliminates some of the gauge freedom of applying arbitrary unitaries to $A'B'$.

It is interesting to consider if this geometrically optimal purification is, in fact, also the minimal purification of the given density matrix. In other words, a natural conjecture to make is that the particular purification obtained using the surface-state correspondence achieves the infimum for multipartite $E_P$ in general, i.e., that
\begin{equation}
E_P(A_1 : \ldots : A_n) = \frac{1}{n} \left[ S(AA_1^\prime) + \cdots + S(AA_n^\prime) \right] .
\end{equation}
This is studied in more detail in \Ref{Bao:2018zab}.

The proposed identification of Hilbert spaces could have interesting applications for the $\text{complexity} = \text{action}$ proposal \cite{Brown:2015lvg, Brown:2017jil}.
In those contexts, the proposed equivalency is unclear regarding for which state the state complexity should be calculated, in particular with regard to the statement of when increasing complexity should result in a firewall. The above notion of minimal-dimension geometric purification gives a candidate state to which to restrict in the statement of the $\text{complexity} = \text{action}$ conjecture. It furthermore has the benefit of making close contact with cases for which the $\text{complexity} = \text{action}$ conjecture is known to be true, specifically the context of shock waves in wormhole geometries. It is an enticing possibility that one can use this work to help make that conjecture more precise.

Note that the cases in which the entanglement wedge is disconnected are handled well in the surface-state correspondence hybridization with $E_P=E_W$.
Disconnected entanglement wedges correspond to boundary states whose reduced density matrices can to leading order be written as tensor products, of which the factors can be separately purified.
This separable purification also implies that one can act with simultaneous isometries from different (but not disjoint) subfactors of the boundary Hilbert space encoding the complement of the entanglement wedge to the disjoint Ryu-Takayanagi surfaces bounding these entanglement wedges separately, yielding the correct $A_i'$s on the respective Ryu-Takayanagi surfaces bounding each portion of the entanglement wedge.

\section{Discussion}
\label{sec:disc}

The entanglement of purification, $E_P$, is conjectured to correspond holographically to the area, $E_W$, of a bulk-anchored minimal surface that connects boundary-anchored minimal surfaces.
We examined the multipartite construction in the first part of this article, which led us to a new class of inequalities relating $n$-partite $E_W$s to sums of lower-partite $E_W$s, as was shown in \Sec{subsec:multi_ineqs}.
Because the geometric description was so crucial in deriving these inequalities, it is unlikely that they extend to $E_P$ inequalities for all states, assuming that $E_P = E_W$.
Rather, they are at least guaranteed to be obeyed by states with dual holographic geometries, similarly to how certain entropy inequalities, such as monogamy of mutual information, hold for such states \cite{Hayden:2011ag}.
That a given CFT\textsubscript{2} state satisfy these $E_P$ inequalities is therefore a necessary condition for it to have a dual holographic geometry.
We expect that these inequalities would be useful in the construction of an entanglement-of-purification cone for holographic states, analogous to the holographic entropy cone describing entanglement entropies \cite{Bao:2015bfa}.
We also expect our inequalities to generalize straightforwardly to dimensions higher than $2+1$; however, we cannot draw any immediate conclusions without further analysis, since the $E_W$ surface bulk-anchoring condition and the assessment of circumscription are somewhat more subtle in higher dimensions.

In the second part of this article, we observed that the calculation of bipartite $E_W$ in AdS\textsubscript{3} can be mapped to a holographic entanglement entropy calculation in a multiboundary wormhole geometry. In short, modding out by isometries maps entanglement wedge cross sections in the Poincar\'e disk to wormhole throats in the resulting wormhole geometry, which are the Ryu-Takayanagi surfaces of entire boundaries.
This relationship strongly hints at the possibility that the correct way to interpret, from an information-theoretic point of view, the procedure of constructing wormhole geometries from vacuum AdS is as some specific optimal purification procedure of, e.g.,  $\rho_{AB}$ to $\rho_{\tilde{A}\tilde{B}}$ in the two-boundary case.
In this case, the CFTs connected by the wormhole geometry would then precisely be the optimal purification of boundary subregions of the original CFT.

In general, minimal surface identifications through isometric quotients can result in subdominant saddle solutions; the bulk solution may prefer to be two or more disconnected asymptotically AdS\textsubscript{3} geometries, for example when the minimal separation between the identified surfaces becomes too small. Those wormhole geometries associated with nontrivial entanglement of purification will, however, remain the dominant geometry, as they always correspond to the phase where the identification region is the connected entanglement wedge, ensuring by definition that the dominant saddle is a wormhole geometry with nonzero throat size.

If $E_P = E_W$, one advantage of the wormhole proposal is that it is calculationally easier to evaluate entanglement entropies than entanglements of purification.
The proposal also lets one convert $E_P$ inequalities into entropy inequalities and vice versa.
As was demonstrated in \Sec{sec:interpretingEP}, we can both reproduce known $E_P$ inequalities and derive new inequalities (for both $E_P$ and $S$) via this identification.

A limitation of the wormhole construction is that entanglement entropy inequalities in the wormhole geometry only convert to $E_P$ inequalities for empty AdS\textsubscript{3}.
Nevertheless, the basic idea behind our construction would apply to any pair of wormhole geometry and parent asymptotically AdS\textsubscript{3} spacetime that possesses isometries that allow for a smooth quotient identification.
In particular, one could imagine building up such parent geometries by locally perturbing a slice of empty AdS\textsubscript{3} in the following schematic way.
Given an entanglement wedge, which serves as the fundamental domain for the wormhole construction, the procedure would be to act with a local perturbation inside the fundamental domain, as well as at the same location in all images of the fundamental domain in the Poincar\'e disk.
A quotient of the perturbed disk by the original isometries then presumably yields a perturbed wormhole geometry.

An important remark that we made at the beginning of \Sec{sec:WHviaOrbifolds} that is worth revisiting here is that our arguments were purely geometric; at no point did we invoke any CFT principles to corroborate our findings about $E_W$ inequalities and the relationship between $E_W$ and wormhole throats.
In particular, regarding the latter finding, a high-priority question is whether one can show that $E_P$ in the parent CFT state is equal to the entanglement entropy of the reduced state on a collection of CFT boundaries, where these boundary CFTs are obtained directly from a quotient in the original CFT.
A purely field-theoretic analysis would not only firmly substantiate our findings, but also lend further credence to the holographic $E_P = E_W$ conjecture more generally.

This research leaves many promising avenues for future investigation.
It would be interesting to consider the extension of our results to dimensions higher than $2+1$.
In higher-dimensional gravity, the special symmetries that allow for simple identification of surfaces in AdS\textsubscript{3} to construct wormholes do not appear, and thus the story is more complex.
One could, of course, simply construct multi-boundary black brane solutions, but this does not appear to offer any more insight than the three-dimensional case.
Nevertheless, it is an interesting open question to ask if there are any higher-dimensional objects that, via some transformation of the bulk geometry possibly generalizing the three-dimensional identification program, would transform the $E_W$ surfaces once again into bulk minimal surfaces that calculate the entanglement entropy in some boundary region of a CFT.

Finally, in a similar vein to field-theoretic analysis, an interesting further direction is to understand more precisely the relationship between the optimal $A_iA'_i$ state that calculates entanglement of purification, the $A_i A'_i$ state obtained from the surface/state correspondence, and the wormhole boundary state $\tilde A_i$.
The most straightforward possibility, which is further strongly motivated by information-theoretic arguments, is that these are one and the same.
Nonetheless, having a concrete mathematical implementation of the maps among these states would be illuminating for our understanding of the relationship in the entanglement structure of holographic states with and without wormholes.

\vspace{5mm}

\begin{center} 
 {\bf Acknowledgments}
 \end{center}
 \noindent 
We thank Raphael Bousso, Tom Faulkner, Illan Halpern, Veronika Hubeny, Alex Maloney, and Jason Pollack for useful discussions and comments. N.B. is supported by the National Science Foundation under grant number 82248-13067-44-PHPXH and by the Department of Energy under grant number DE-SC0019380. A.C.-D. was supported by a Beatrice and Sai-Wai Fu Graduate Fellowship in Physics and by the Gordon and Betty Moore Foundation through Grant 776 to the Caltech Moore Center for Theoretical Cosmology and Physics for the majority of the work. A.C.-D. is currently supported in part by the KU Leuven C1 grant ZKD1118 C16/16/005, the National Science Foundation of Belgium (FWO) grant G.001.12 Odysseus, and by the European Research Council grant no. ERC-2013-CoG 616732 HoloQosmos. G.N.R. is supported by the Miller Institute for Basic Research in Science at the University of California, Berkeley.

\bibliographystyle{utphys-modified}
\bibliography{Allrefs_inspirehep}

\end{document}